\documentclass[12pt]{article}

\usepackage{graphicx}
\usepackage{times}
\usepackage{color}

\usepackage{ulem}
\usepackage{lineno}
\usepackage{hyperref}
\usepackage{multirow}
\usepackage{makecell}
\usepackage{siunitx}
\usepackage{adjustbox}
\usepackage{float}
\usepackage{amsmath, amssymb}
\usepackage{soul}

\DeclareSIUnit\cells{cells}

\newcommand{\pilH}[1][]{$\Delta${\it pilH} }

\title{Evidence of robust, universal conformal invariance in living biological matter} 

\author
{Benjamin H. Andersen$^{1}$, Francisco M. R.  Safara$^{2,3}$,\\ Valeriia Grudtsyna$^{1}$, Oliver J. Meacock$^{4,5}$,\\ Simon G. Andersen $^{3}$, William M. Durham$^{5\ast}$, Nuno A. M. Araujo$^{2,3\ast}$\\ Amin Doostmohammadi$^{1\ast}$\\
\\
\normalsize{$^{1}$Niels Bohr Institute, University of Copenhagen, Copenhagen, Denmark}\\
\normalsize{$^{2}$Departamento de Física, Faculdade de Ciências, Universidade de Lisboa, Lisboa, Portugal}\\
\normalsize{$^{3}$Centro de Física Teórica e Computacional, Faculdade de Ciências,}\\ \normalsize{Universidade de Lisboa, Lisboa, Portugal}\\
\normalsize{$^{4}$Department of Fundamental Microbiology, University of Lausanne, Lausanne, Switzerland}\\
\normalsize{$^{5}$Department of Physics and Astronomy, University of Sheffield, Sheffield, United Kingdom}\\
\normalsize{$^\ast$To whom correspondence should be addressed:}\\
\normalsize{\href{mailto:w.m.durham@sheffield.ac.uk}{w.m.durham@sheffield.ac.uk }, \href{mailto:nmaraujo@fc.ul.pt}{nmaraujo@fc.ul.pt }, \href{mailto:doostmohammadi@nbi.ku.dk}{doostmohammadi@nbi.ku.dk }}
}

\date{}

\begin{document} 




\maketitle 

\begin{abstract}
Collective cellular movement plays a crucial role in many processes fundamental to health, including development, reproduction, infection, wound healing, and cancer. The emergent dynamics that arise in these systems are typically thought to depend on how cells interact with one another and the mechanisms used to drive motility, both of which exhibit remarkable diversity across different biological systems. Here, we report experimental evidence of a universal feature in the patterns of flow that spontaneously emerges in groups of collectively moving cells. Specifically, we demonstrate that the flows generated by collectively moving dog kidney cells, human breast cancer cells, and by two different strains of pathogenic bacteria, all exhibit conformal invariance. Remarkably, not only do our results show that all of these very different systems display robust conformal invariance, but we also discovered that the precise form of the invariance in all four systems is described by the Schramm-Loewner Evolution (SLE), and belongs to the percolation universality class. A continuum model of active matter can recapitulate both the observed conformal invariance and SLE form found in experiments. The presence of universal conformal invariance reveals that the macroscopic features of living biological matter exhibit universal translational, rotational, and scale symmetries that are independent of the microscopic properties of its constituents. Our results show that the patterns of flows generated by diverse cellular systems are highly conserved and that biological systems can unexpectedly be used to experimentally test predictions from the theories for conformally invariant structures.\\

\end{abstract}

Understanding the collective movement of large populations, and how it arises from its constituents, is a central problem in biology, ecology, material science and physics~\cite{bechinger_active_2016,julicher_hydrodynamic_2018,friedl2009collective,zhang2021autonomous}.  In these living systems, work is produced at the level of an individual constituent, and this `activity' is translated into patterns of collective motion at larger length scales through interactions between them~\cite{marchetti_hydrodynamics_2013,bechinger_active_2016}. However, many of the processes involved in collective movement, including the mechanisms that individual constituents use to propel themselves, the processes that give rise to interactions, and the behavioural responses to stimuli, are incredibly diverse in different biological systems and are often difficult to decode ~\cite{wadhwa2022bacterial,trepat2018mesoscale}. While many different models have been proposed to reproduce the specific pattern of collective movement made by particular organisms, we lack a general unifying theory or set of principles that unite the collective movement observed across distinct biological systems. 

In contrast, the study of the complex interactions between the components that make up certain inanimate materials, like metals and alloys, has led to the discovery of universal behavior near the so-called critical regimes. In these conditions, the global macroscopic properties no longer depend on the specific properties of the individual constituents, but rather exhibit ``universal" behavior ~\cite{hohenberg_theory_1977}. The principles that give rise to this universality in inanimate materials have been described using the framework of conformal field theory~\cite{belavin1984infinite,cardy1984conformal}, which predicts how shapes and angles of structures will be locally conserved across different systems, but not necessarily their length scales or curvatures. 
While the techniques used to describe conformally invariant structures have long been used to make theoretical predictions in statistical mechanics and condensed matter physics~\cite{belavin1984infinite,cardy1984conformal} and to establish the universality of critical phenomena (for example, using numerical studies of turbulence~\cite{Bernard2006,bernard2007inverse,puggioni2020conformal} and rigidity percolation~\cite{javerzat2023evidences,javerzat2024schramm}), direct experimental observation of conformal invariance and robust universal critical behavior in living matter remains elusive.  

In this paper, we experimentally demonstrate that the patterns of collective movement observed in different types of living matter exhibit universal characteristics that transcend the particular properties of the cells from which they are composed. We show that vastly different systems, including colonies of pathogenic bacteria, groups of collectively moving dog kidney cells, and human breast cancer cells, all spontaneously generate flows that exhibit a universal conformal invariance that can be described by the percolation universality class. 
This finding suggests that collective cellular movement, which plays an important role in many biological systems~\cite{friedl2009collective,guillamat2022integer,prasad2022alcanivorax}, could potentially serve as a fundamental test bed for theories that are based on conformal symmetry.

\begin{figure}[ht!]
\centering
\includegraphics[scale=0.5]{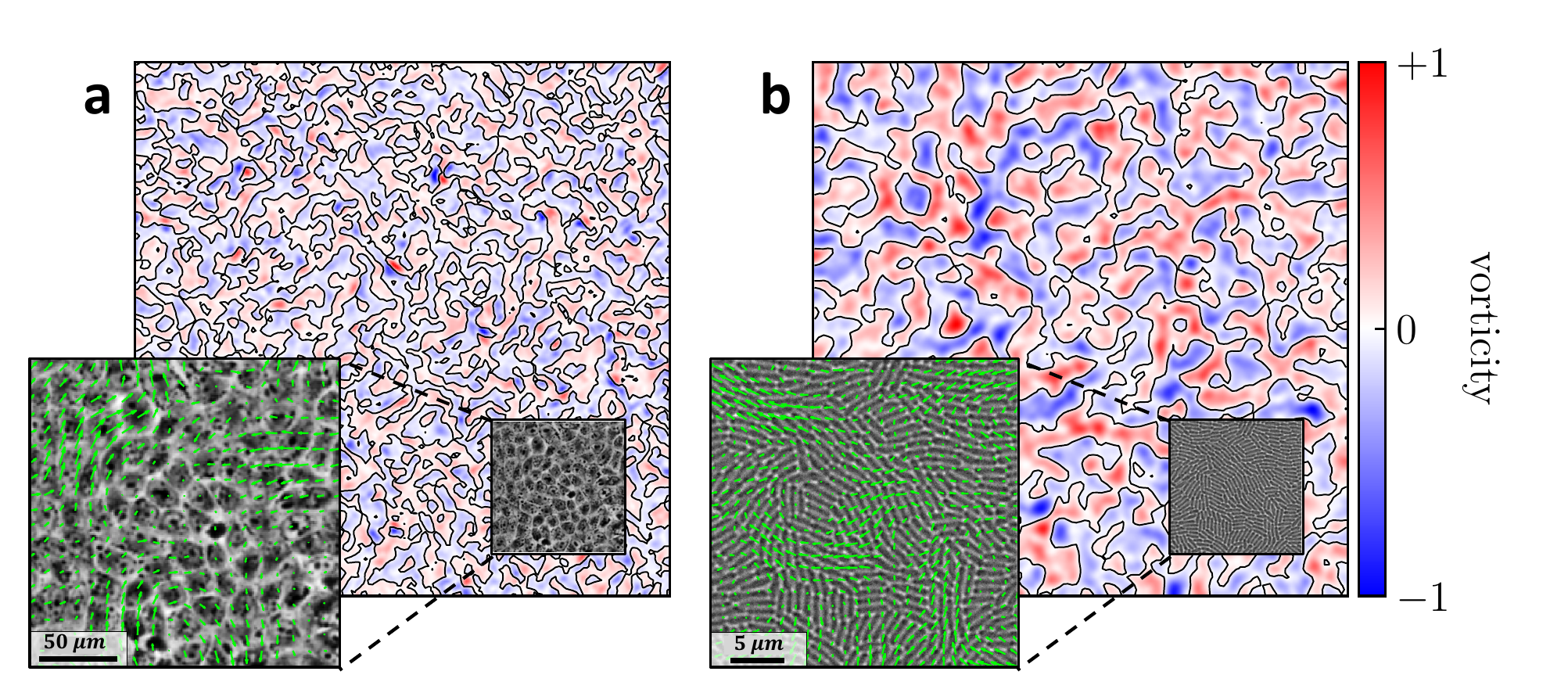}
\caption{\textbf{High-resolution measurements of the coherent flows from collectives of eukaryotic and prokaryotic cells.} Representative velocity and vorticity fields observed in monolayers of (\textbf{a}) eukaryotic Madin-Derby Canine Kidney (MDCK) cells, and (\textbf{b}) prokaryotic wild-type {\it Pseudomonas aeruginosa} cells. The colormap shows the local vorticity and zero-vorticity contours are marked with black lines. The vorticity is normalised by its maximum value. Insets show a subset of the cells within a single field of view, which have been overlaid with green arrows showing the local velocity. Here we have quantified movement using single cell tracking (PTV), but we have also verified our results using particle image velocity (PIV) (\textit{Materials and Methods}). 
}
\label{fig:1}       
\end{figure}
We made high-resolution measurements of monolayers composed of four different cellular genotypes, including both prokaryotes and eukaryotes, to resolve whether we could identify common features in their collective motility. 
For prokaryotes, we studied the opportunistic pathogen {\it Pseudomonas aeruginosa}, which uses tiny grappling hooks called pili to pull themselves along solid surfaces, a process known as twitching motility~\cite{meacock_bacteria_2021}. We considered two different strains of this rod-shaped bacteria -  wild type (WT) PAO1 and a deletion mutant \pilH lacking one of the response regulators in the Pil-Chp system, which causes it to become hyperpiliated, move faster, and form longer cells than its parental WT ~\cite{meacock_bacteria_2021, oliveira2016, bertrand2010genetic}. For the eukaryotic cells, we considered the commonly studied Madine-Darby-Canine-Kidney (MDCK) cells~\cite{saw2017}, and aggressive human breast cancer cells (MCF-7)~\cite{balasubramaniamActive2021}. Each of these genotypes forms monolayers through in-situ growth and exhibits two-dimensional collective patterns of motion when they reach confluence. Vortical flow structures, a characteristic feature of the disordered flows observed in wide diversity of different systems~\cite{pismen1999vortices}, are observed in all four of the cellular genotypes investigated here (e.g. see Fig.~\ref{fig:1}). Each vortex either exhibits clockwise or anti-clockwise rotation and the line that sits at the boundary between flows that rotate in opposite directions, the zero-vorticity contour, provides a measure of the underlying structure of the flow. 

\begin{figure}[ht!]
\centering
\includegraphics[scale=.95]{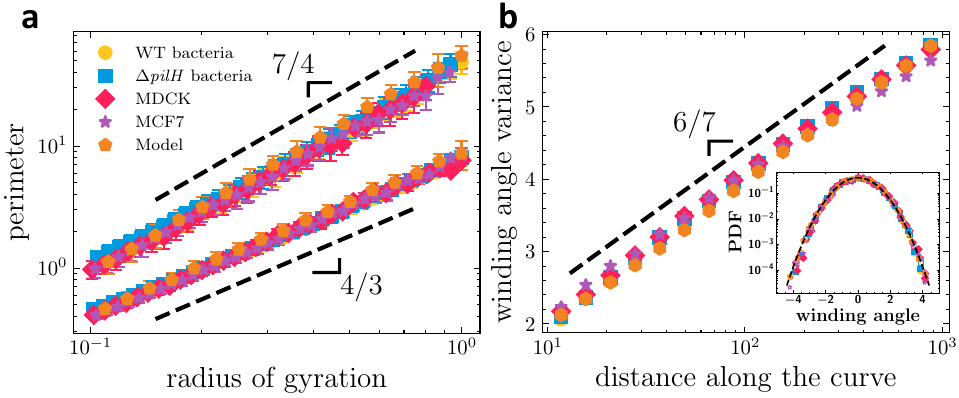}
\caption{\textbf{Vorticity contours from four distinct cellular systems exhibit the same patterns of scale and conformal invariance, which is recapitulated using a continuum model of active fluids.} \textbf{(a)} The perimeter of contours as a function of their radius of gyration for two prokaryotic and two eukaryotic genotypes, including wild-type {\it P. aeruginosa} bacteria (yellow circles) and a hyperpilated \pilH {\it P. aeruginosa} mutant (blue squares) that individually move faster, Madine-Darby Canine Kidney cells (red diamonds), and MCF-7 human breast cancer cells (purple stars). Here we separately analysed the complete perimeter and accessible external perimeter of the contours (SI Fig. S2) - we found that the experimental data for all four genotypes collapsed onto lines with slopes of approximately $7/4$ and $4/3$ respectively for the two different perimeter measurements. The flow fields produced by a numerical model of active fluids (\textit{Materials and Methods}) generated vorticity contours with a power-law dependency in close agreement with that observed in experiments. The radius of gyration is normalised by the maximum system size (\textit{Materials and Methods}).
(\textbf{b}) The  variance in the distribution of the winding angle, plotted here as a function of distance along the curve for the four experimental systems and numerical model, all exhibit the same logarithmic scaling with a slope of $6/7$ (dashed line). (\textbf{b}; inset) In addition, the distribution of winding angles for a fixed distance along the contour is closely approximated by a Gaussian (dashed line).  Both findings are consistent with that predicted for conformally invariant curves, which exhibit the same fractal dimension that we obtained for our data in Fig.~\ref{fig:2}a. The dashed lines correspond to the slope $6/7$ and a standard Gaussian distribution, respectively. In (\textbf{b}; inset) the winding angles
are obtained for segments of contours of length $64$ (filled symbols) and $512$ (empty symbols) and measured relative to the average angle of the contour. Results are averages over different samples and error bars represent the standard deviation (see \textit{Materials and Methods}).}
\label{fig:2}    
\end{figure}

To compare how the flow structure varies across the four different experimental systems, we first measure the fractal dimension of the vorticity contours by plotting the perimeter of closed contours as a function of their radius of gyration. Surprisingly, without any fitting, special scaling, or free parameters, the results for all four different experiments collapse on the same line and exhibit the same power-law behavior (Fig.~\ref{fig:2}a). This provides concrete evidence of scale invariance and indicates that the flows generated by these diverse cellular systems share the same generic features. Interestingly, the slope of the perimeter-gyration radius plot, or fractal dimension, is $D=7/4$ for the complete perimeter and $D_\ast=4/3$ for the accessible external perimeter (Fig. S2) and satisfies the duality relation $4(D-1)(D_\ast-1)=1$, conjectured for conformally invariant curves~\cite{duplantier2000conformally}. This finding suggests that these biological flow structures, in addition to being scale invariant, could exhibit much richer conformal symmetries~\cite{duplantier2003higher}.

To test whether our experimental data exhibits conformal invariance, we calculated the winding angle of the vorticity contours across the four different experimental systems. The winding angle is defined as the angle between two points on a contour that are separated by a given distance measured along the contour (\textit{Materials and Methods}). For conformally invariant curves ({\it i}) the winding angles are Gaussian distributed and ({\it ii}) the variance in the distribution of winding angles increases logarithmically with the length of the curve ~\cite{duplantier_wang}. Our experimental data is in close agreement with both predictions for conformal invariance – with both metrics collapsing the data from the vorticity contours of the four cellular systems onto the same line (Fig.~\ref{fig:2}b and Fig. S4). Moreover, the rate at which the variance of the winding angle increases with the logarithm of the length is predicted to scale as $\alpha= 6/7 = 2(D-1)/D$ for conformally invariant curves~\cite{duplantier_wang}. Thus, for the fractal dimension of $D = 7/4$ we measured in Fig.~\ref{fig:2}a, we would predict that $\alpha = 6/7$, which is supported by our direct measurements of the variance (Fig.~\ref{fig:2}b). 

\begin{figure}[ht!]
\centering
\includegraphics[scale=0.95]{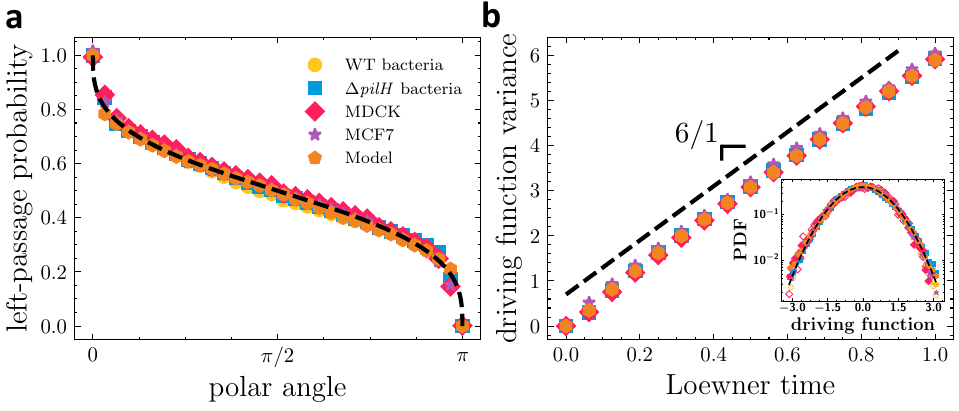}
\caption{\textbf{Resolving the underlying universality class of vorticity contours using two independent methods.} \textbf{(a)} The left-passage probability is defined as the probability that a point in space is on the right-hand side of the contour for a given polar angle. Data from all four cellular genotypes and the results from the numerical model are in close agreement with Schramm’s formula for $\kappa=6$ (dashed black line, ~\cite{Schramm_formula}). \textbf{(b)} Time-dependence of the variance of driving function obtained from a unique conformal slit map~\cite{Kennedy_2009}. The dashed black line shows the result for one-dimensional Brownian motion with $\kappa = 6$. \textbf{(b, inset)} Probability distribution of the driving function, rescaled by $\kappa t$, where $t$ is the Loewner time.  Here the data at two different times $t = 0.25$ and $t = 0.75$ is shown, which collapses onto the same curve. For further details see \textit{Materials and Methods}.}
\label{fig:4}
\end{figure}

Our results strongly suggest that the flows spontaneously generated by diverse cellular genotypes exhibit robust conformal invariance. We next sought to ascertain we could resolve which universality class these biological flows belong to. One of the central mathematical breakthroughs of the last few decades was to demonstrate that certain systems with conformal invariance and domain Markov property, can be described, in the scaling limit of interfaces, by a family of planar curves defined by a single parameter $\kappa$. This formalism is known as the Schramm-Loewner Evolution (SLE)~\cite{Schramm_2000,cardy2005sle} and the value of the $\kappa$ distinguishes different fundamental statistical mechanics models at criticality and thus resolves the universality class that a system belongs to ~\cite{Kennedy_2002,SMIRNOV,domain_walls}. To determine if the vorticity contours in the cellular systems are SLE curves, we extracted the $\kappa$ parameter from the four experimental systems. We used two distinct and independent methods~\cite{daryaei2012watersheds}: ({\it i}) directly calculating the driving function~\cite{Kennedy_2009}, and ({\it ii}) measuring the left-passage probability, comparing both to analytic predictions for the SLE~\cite{Schramm_formula} (\textit{Materials and Methods}). Both methods consistently yielded $\kappa = 6$, for all four cellular genotypes  (Fig.~\ref{fig:4}). The value of $\kappa = 6$ is also consistent with the estimated fractal dimension, ($D$, Fig.~\ref{fig:2}a), which for SLE is related to $\kappa$, as $D = 1 + \kappa/8$~\cite{beffara_2008}. This particular value of $\kappa$ has an important physical meaning, as it has been uniquely proven for $\kappa = 6$ that SLE curves correspond to the contours of critical percolation clusters and have the locality property (such that the properties only depend on the immediate neighborhood)~\cite{SMIRNOV,lawler2011values}. As such, our analyses reveal that the vorticity contours found in the four different cellular systems are not only conformally invariant, but they also all fall into the universality class of percolation.

Our experimental results indicate that diverse cellular types collectively generate flows with remarkably similar patterns of scale and conformal invariance, implying that the physical mechanisms that underlie the flow structures are highly conserved.  
While many different physical models of active matter have been developed to approximate specific types of cells and the processes unique to them~\cite{marchetti_hydrodynamics_2013,julicher_hydrodynamic_2018,shaebani_computational_2020}, we tested whether a generic model could recapitulate our experimental observations. We used a simple continuum model in which a nematic order parameter (corresponding to cell orientation) was coupled to an incompressible velocity field (\textit{see Materials and Methods} for further details). The two main parameters are the activity $\zeta$, represented by active stress generation in the velocity equation, and elasticity, represented by the elastic constant $K$ that penalizes deformations. Dimensional analysis of the governing equations yields a characteristic length scale of $K/\zeta$, which defines the fundamental length scale of the flow. We find that the vorticity contours of this minimal model recapitulate each of the  measurements observed in our experimental systems, including the fractal dimension $D=7/4$ (Fig.~\ref{fig:2}a), winding angle scaling $\alpha=6/7$ (Fig.~\ref{fig:2}b), and the scaling of the driving function and left-passage probability $\kappa=6$ (Fig.~\ref{fig:4} and Table~\ref{table : results}).  While differences in cell morphology, intercellular adhesion, mechanotransduction, and the mechanisms that give rise to local alignment can all affect patterns of collective motility~\cite{wensink_meso-scale_2012,ladoux_mechanobiology_2017,trepat2018mesoscale}, the results of our continuum model imply that such idiosyncratic characteristics do not materially influence scale and conformally invariant flow patterns, but rather they are a generic feature of collective cellular flows.

The observed scaling of the vorticity contours from both experiments and model are compatible with the Schramm-Loewner evolution (SLE) with $\kappa = 6$, for more than two decades in range and this was confirmed using two independent methods (Fig.~\ref{fig:4} and Table~\ref{table : results}). Remarkably, this finding demonstrates that, although the collective cellular motility spontaneously generates patterns of flow with lengthscales much larger than that of individual cells (Fig.~\ref{fig:1})~\cite{wensink_meso-scale_2012,friedl2009collective}, the associated vorticity contours are local and fall into the same universality class as those from random percolation~\cite{SMIRNOV}.
Moreover, the collective cellular motion we studied here is driven far from equilibrium by the motility of individual cells that continuously inject energy into the system at small scales.
The observation of conformal invariance in collective cellular flows that are continuously driven far from thermodynamic equilibrium presents both challenges and new opportunities for the development of non-equilibrium conformal field theories~\cite{bernard2016conformal}.
\begin{table}[h!]
\centering
\begin{adjustbox}{max width=\textwidth}
\begin{tabular}{|ccc|c|cccc|c|}
\hline
\multicolumn{3}{|c|}{\multirow{3}{*}{}}
& \multirow{3}{*}{\bf\makecell{continuum \\ model}} & \multicolumn{4}{c|}{\bf experimental data} & \multirow{3}{*}{\bf\makecell{percolation \\ universality \\ class}} \\ \cline{5-8}
\multicolumn{3}{|c|}{} & & \multicolumn{2}{c|}{bacterial cells} & \multicolumn{2}{c|}{eukaryotic cells} & \\ 
\cline{5-8}\multicolumn{3}{|c|}{} & & \multicolumn{1}{c|}{WT} & \multicolumn{1}{c|}{\pilH} & \multicolumn{1}{c|}{MDCK} & MCF7 & \\ \hline
\multicolumn{1}{|c|}{\bf scale invariance} & \multicolumn{1}{c|}{fractal dimension} & $D$ & $1.75 \pm 0.01$ & \multicolumn{1}{c|}{$1.72 \pm 0.02$} & \multicolumn{1}{c|}{$1.72 \pm 0.04$} & \multicolumn{1}{c|}{$1.74 \pm 0.02$} & $1.74 \pm 0.03$ & $7/4 = 1.75$ \\ 
 \hline
\multicolumn{1}{|c|}{\bf conformal invariance} & \multicolumn{1}{c|}{winding angle} & $\alpha$ & $0.853 \pm 0.005$ & \multicolumn{1}{c|}{ $0.86 \pm 0.01$} & \multicolumn{1}{c|}{$0.87 \pm 0.01$} & \multicolumn{1}{c|}{$0.85 \pm 0.01$} & $0.87 \pm 0.02$ & $6/7 \approx 0.857$ \\ \hline
\multicolumn{1}{|c|}{\multirow{2}{*}{\bf\makecell{Schramm-Loewner \\ Evolution}}} & \multicolumn{1}{c|}{left-passage probability} & $\kappa$  & $6.02 \pm 0.02$ & \multicolumn{1}{c|}{$5.97 \pm 0.05$} & \multicolumn{1}{c|}{$5.96 \pm 0.05$} & \multicolumn{1}{c|}{$5.95 \pm 0.03$}  & $5.95 \pm 0.06$ & $6$ \\ \cline{2-9} 
\multicolumn{1}{|c|}{} & \multicolumn{1}{c|}{driving function} & $\kappa$ & $5.96 \pm 0.05$ & \multicolumn{1}{c|}{$6.03 \pm 0.06$} & \multicolumn{1}{c|}{$5.96 \pm 0.06$} & \multicolumn{1}{c|}{$5.98 \pm 0.04$} & $5.93 \pm 0.04$ & $6$ \\ \hline
\end{tabular}
\end{adjustbox}
\caption{\textbf{Measurements of the fractal dimension, winding angle, left-passage probability, and driving function of the four different cellular genotypes and numerical simulations.} The values are calculated from the velocity fields obtained from single cell tracking ({\it Materials and Methods}) and the errors represent standard deviation about the mean.}
\label{table : results}
\end{table}

These results suggest that the theories used to describe conformally invariant structures might have a much broader range of applications than previously anticipated. While collective movement is observed in diverse biological systems, that observed in microscopic cellular systems is particularly amenable to experimental analysis because in-situ cell division rapidly gives rise to large genetically identical populations, two-dimensional movement of monolayers of cells can be readily imaged, and the environmental conditions can be carefully controlled. Similar to the collective cellular motility studied here, many different kinds of living systems are formed of strongly interacting components driven far from thermal equilibrium and exhibit complex vortical patterns, including subcellular flows~\cite{goldstein2015physical,stein2021swirling}, synthetic active material~\cite{palacci2013living,bechinger_active_2016,han2020emergence}, animal swarms~\cite{cavagna_bird_2014,delcourt2016collective}, and {\it in-vitro} reconstitutions of cytoskeletal transport systems~\cite{schaller2010polar,sanchez_spontaneous_2012,sumino2012large}. 
In addition, emergent vortical structures also shape many important processes in biology including cell differentiation~\cite{guillamat2022integer}, cartilage regeneration~\cite{makhija2022topological}, embryogenesis~\cite{Smith2019}, signaling waves that propagate along cell membranes~\cite{tan2020topological} and between cells~\cite{mathijssen2019collective}, as well as vortical waves associated with cardiac arrhythmia~\cite{christoph2018electromechanical} and spiral-like patterns of brain activity linked to cognitive processing~\cite{xu2023interacting}. We speculate that such biological processes might not only serve as novel test bed to validate predictions based on conformal symmetry, but this robust symmetry might also lead to the development of new analytical techniques to identify the fundamental mechanisms that give rise to both function and dysfunction in complex biological systems.

\bibliography{references}
\bibliographystyle{ieeetr}

\section*{Acknowledgments}
We thank Guido Boffetta for comments on a previous version of this manuscript and Ramin Golestanian for helpful discussions.

\section*{Funding}
This work was supported by the Novo Nordisk Foundation (grant no. NNF18SA0035142 and NERD grant no. NNF21OC0068687) (to AD), Villum Fonden Grant no. 29476 (to AD), the European Union via the ERC-Starting Grant PhysCoMeT, grant no. 101041418 (to AD), the Portuguese Foundation for Science and Technology (FCT) under Contracts no. EXPL/FIS-MAC/0406/2021, UIDB/00618/2020, and UIDP/00618/2020 (to NA), a Biotechnology and Biological Sciences Research Council (BBSRC) New Investigator grant (grant no. BB/R018383/1) (to WMD), and a Human Frontier Science Program grant (grant no. RGY0080/2021) (to WMD).

\section*{Authors contributions}
A.D. and N.A.M.A. designed the project. B.H.A. and F.S. performed analyses of scale and conformal invariance as well as SLE measurements on the experimental data. B.H.A. implemented the model, analysed data and prepared figures. V.G. performed the experiments and conducted particle image velocimetry measurements on MDCK and MCF-7 cells. S.G.A. performed single cell tracking analyses for MDCK and MCF-7 monolayers. O.J.M. performed the experiments on bacterial cells and conducted particle image velocimetry and single cell tracking analyses on bacterial monolayers. W.M.D., N.A.M.A. and A.D. all contributed to the design of experiments and models, as well as to the interpretation of results. A.D. prepared the first draft. A.D. and W.M.D. wrote the paper with input from N.A.M.A. This collaborative effort was led by A.D.

\end{document}